\documentclass[12pt,a4paper]{article}
\usepackage{amsmath,amssymb,amsthm,mathrsfs,amsfonts,dsfont} 
\usepackage{color,caption}
\usepackage{graphicx}
\usepackage{float}

 \newcommand{\be}{\begin{equation}}
 \newcommand{\ee}{\end{equation}}
 \newcommand{\ba}{\begin{eqnarray}}
 \newcommand{\ea}{\end{eqnarray}}

\newcommand{\no}{\nonumber\\}

\begin{document}
			
	\begin{center}
		{\Large \bf On the gravity dual of the light-front vacuum} \\
		\vskip .5cm
		Blai Garolera\\
		\centerline{\it Escuela de F{\'\i}sica,
			Universidad de Costa Rica}
		\centerline{\it 11501-2060 San Jos\'e,
			Costa Rica}
		\bigskip
		\verb"blai.garolera@ucr.ac.cr"
		
	\end{center}
	\bigskip\bigskip
	\begin{abstract}
		Building on a previous conjecture \cite{Cvetic:1998jf,Brecher:2000pa}, we argue that the holographic dual of the light-front vacuum state of a superconformal field theory quantized in the front form of dynamics contains the Kaigorodov spacetime, which is nothing but a pp-wave propagating in AdS. Evidence in favor of this conjecture is presented. In particular we verify the matching of global symmetries and discuss the contribution of the zero mode sector in both sides of the correspondence.
		\bigskip
	\end{abstract}

	\section{Introduction}
	Light-front quantization\footnote{For a comprehensive review see for example \cite{Brodsky:1997de, Heinzl:1998kz, Heinzl:2000ht}.} (LFQ) of quantum field theories refers to a specific choice for the hypersurface where we decide to specify the initial conditions and canonical commutation relations for a given dynamical problem and provides us with a powerful alternative to ordinary equal-time quantization. While the usual option in field theory is to impose equal-time commutation relations at a particular instant of time $ct=x^0=0$, in light-front quantization we impose such commutators on a light-front $x^+=\frac{1}{\sqrt{2}}(ct+x^p)=0$, that is, on the hypersurface generated by a plane light wave traveling in the negative $x^p$ direction. The light-like coordinate $x^+$ plays now the role of time and $P^-$, the generator of translations in light-front time, plays accordingly the role of the (light-front) Hamiltonian. \\
	\indent  One of the key features of LFQ stems from the fact that when we rewrite the $(p+1)$-dimensional Poincar\'e algebra in light-front coordinates it appears a subalgebra isomorphic to the Lie algebra of the $(p-1)$-dimensional Galilei group in the directions transverse to the light-front. In turn, such a Galilean covariance implies that the light-front kinematics will partly show a non-relativistic	behavior. This unique peculiarity is specially profitable when we deal with a system of many interacting particles and, in particular, it can lead to a relativistic description of bound systems in terms of quantum-mechanical wave functions, that is, solutions of a field-theoretic analog of the Schr\"odinger equation of non-relativistic quantum mechanics.  On top of that, the light-front wave functions describing the states are each frame-independent and in general it can be seen that light-front bare states are closer to the partons that are observed to make up hadrons than the equal-time bare states. Because of these particular features, light-front quantization has been considered for many years as a very promising approach to solve the non-perturbative dynamics of quantum chromodynamics (QCD) and the ideal framework for a unification of two fundamentally different pictures of the hadronic matter, namely the constituent quark model and the non-abelian quantum field theory that is QCD. \\
	\indent  Another fundamental aspect of LFQ is the fact that the physical light-front vacuum (LFV) is simple, or even trivial. By trivial we mean that the LFV coincides with the Fock space vacuum and at the same time it is an eigenstate of the full interacting light-front Hamiltonian. This of course is a significant advantage for LFQ compared to the standard equal-time quantization. For instance, with a trivial physical vacuum, the solution of the hadronic eigenvalue problem for the light-front Hamiltonian of QCD can then focus only on the massive states, unlike equal-time quantization where the vacuum state itself is non-trivial and must be computed as well. However, such a trivial vacuum can be realized only if we can remove the so-called zero modes (that is, modes with zero longitudinal momentum $P^+$) out of the physical Fock space. This is a subtle issue, referred in the literature as the ``zero mode problem'' \cite{Maskawa:1975ky, Yamawaki:1998cy}. Nevertheless, in some cases it can be shown that zero modes are not dynamical degrees of freedom but instead satisfy a constraint equation through which the zero modes become dependent on the other dynamical modes. If this is the case, the zero modes can be consistently removed from the physical Fock space by solving the zero mode constraint equation, thus establishing the triviality of the LFV. Such a trivial vacuum, on the other hand, must be compatible with the usual picture of complicated non-perturbative vacuum structure in equal-time quantization responsable for confinement, spontaneous symmetry breaking, condensates and vacuum polarization, among others.  Thus, when the LFV can be proved to be trivial, the only possibility to realize such phenomena would be through the complicated structure of the zero mode operators. All in all, in order to study the dynamics and phenomena usually associated with the complicated structure of the vacuum state in equal-time quantization, we essentially need to solve the zero mode constraint equation, either perturbatively or non-perturbatively.\\
	\indent  On a different note, over the last two decades the AdS/CFT correspondence \cite{Maldacena:1997re}, together with its many generalizations now referred altogether to as the gauge/string duality \cite{Aharony:1999ti}, have provided a novel approach for studying the strong coupling limit of a large class of non-abelian quantum field theories. Among many applications, there has been recently an interest in exploiting this technology to study properties of the strongly coupled quark-gluon plasma phase of QCD-like theories at non-zero temperature such as transport coefficients, the jet quenching parameter and several other phenomena characteristic of ultra-relativistic heavy ion collisions\footnote{This particular approach is usually known in the literature  by the name of AdS/QCD. For an extensive review see for example \cite{CasalderreySolana:2011us}.}. Although it is perfectly fair to say that the field theories available through the AdS/CFT correspondence still differ significantly from quantum chromodynamics, many of the results achieved through holography show a surprisingly good agreement with the QCD data and it is generally accepted that the holographic approach may offer insightful analytical approximations to describe the strongly coupled confining dynamics of QCD.\\
	\indent Altogether, a reasonable question arise: can we combine both light-front quantization and the gauge/string holographic duality simultaneously? And if so, do the properties and characteristic features of both approaches benefit each other? Are they compatible? The aim of the present note is precisely this, to take the first step towards the combination of both techniques. More concretely, since the AdS/CFT dictionary establishes that, in a certain regime of the parameter space of the theory, there is a precise duality between a state in the field theory side and a particular geometry or solution of supergravity in the string theory side, the very first question we wanted to answer was precisely to find which could be the gravitational dual of the light-front vacuum state of a specific QFT and how is the zero mode sector implemented holographically. Reinterpreting previous results \cite{Cvetic:1998jf,Brecher:2000pa}, we conjecture that the gravitational dual of the light-front vacuum state of certain $d$-dimensional SCFTs is a pp-wave spacetime of the form of $AdS_{d+1}$ plus a gravitational plane wave propagating parallel to the conformal boundary, the simplest example being the Kaigorodov spacetime \cite{Kaigorodov}. In addition, it is precisely the plane wave which induces a non-vanishing vev of the energy-momentum tensor in the dual field theory, which leads us to propose that the zero mode sector is implemented holographically through the addition of the wave. Furthermore, being this pp-wave an exact solution of supergravity, the profile of the wave is characterized by a function that satisfies a certain differential equation, which in turn may be interpreted as the holographic manifestation of the zero mode constraint equation.\\
	\indent As a final remark, it is worth pointing out that there have been previous attempts in the literature to combine light-front quantization with the gauge/gravity correspondence. The most notable example is Light-Front Holography \cite{Brodsky:2014yha}, which uses the connections and similarities between light-front dynamics, the AdS/CFT correspondence and conformal quantum mechanics. In fact, the present work is aimed to be seen as a ``top-down'' counterpart complementing Light-Front Holography, which is somehow a more ``bottom-up'' approach with a focus on the phenomenology of QCD. \\
	\indent 	The structure of the paper is as follows. In section 2 we review the basics of LFQ, including the separation in kinematical and dynamical generators, the peculiarities of the Poincar\'e and conformal algebras in light-front variables, its Galilei or Schr\"odinger subgroups and the positivity of the kinematical longitudinal momentum operator $P^+$. We also discuss the structure of the light-front vacuum, the construction of the light-front Fock space and the necessity of introducing zero modes. In section 3 we revisit the pp-wave geometries presented in \cite{Cvetic:1998jf,Brecher:2000pa} as well as their holographic computation of the vev of the CFT energy-momentum tensor. Then we argue that is seems more rigorous and accurate to interpret such spacetimes as the dual of a field theory quantized in the light-front and in the light-front vacuum state rather than the gravity dual of a CFT in an infinitely boosted frame and in a certain undetermined state. We conclude by establishing the matching of global bosonic symmetries on both sides of the conjectured duality.
	
	\newpage
	\noindent {\bf Notation and conventions}: Throughout the present text we will be working with a mostly positive metric signature
	 $\eta_{\mu\nu}=\mbox{diag}(-1,+1,\cdots,+1)$, $d=p+1$ denotes the number of spacetime dimensions of the field theory/$p$-brane while $D$ stands for the number of spacetime dimensions in the bulk. The indices $\mu,\nu=0, \cdots, p$ define space-time coordinates, $i,j=1,\cdots, p$ spatial coordinates and $a,b=1,\cdots, p-1$ the coordinates transverse to the light-front. Unless otherwise stated we work with natural units $c = \hbar = 1$.	
	
	\section{Fundamentals of light-front quantization}	
	\subsection{The front form of relativistic dynamics}
	The very essence of light-front quantization (LFQ) first appeared in Paul Dirac's seminal paper \cite{Dirac:1949cp}. According to Dirac's classification, the generators of the Poincar\'e algebra can be separated into \emph{kinematical} and \emph{dynamical}. On the one hand, the kinematical generators act along the hypersurface where the initial conditions are imposed (the Cauchy surface of the problem) and thus leave invariant such a hypersurface. They are independent of the dynamics, don't contain interaction terms and have simple expressions. On the other hand, the dynamical generators (or Hamiltonians, in the sense that they play jointly the role of the single Hamiltonian in non-relativistic quantum mechanics) do modify the Cauchy surface and are the agents responsable for the evolution of the system, i.e. they map one initial surface into another one. Consequently they depend on the different interactions and will exhibit in general a complicated expression.\\
	\indent According to this classification, Dirac introduced three distinct ways to minimally include interactions, the so-called \emph{instant-form}, \emph{point-form} and \emph{front-form} of relativistic dynamics\footnote{Almost three decades later Leutwyler and Stern \cite{Leutwyler:1977vy} found two additional forms of dynamics, for a total of five inequivalent forms, which correspond to the number of possible subgroups of the Poincar\'e group. Nevertheless, such forms present even more dynamical generators than the instant and point forms and thus are not very useful in practice.}. The instant form is the usual one, where the initial surface is $\Sigma_0$, defined by $x^0=0$. In the point-form the initial surface is the hyperboloid $x^\mu x_\mu=-\kappa^2<0,\ x^0>0$. Finally, in the front form of dynamics the initial surface is $\Sigma_+$, the hypersurface generated by the front of a plane light wave propagating in the negative $x^p$-direction, defined by $x^0+x^p=0$. In the instant form there are $\frac{1}{2}p(p+1)$ kinematic generators, the momenta $P^i$ and the angular momenta $M^{ij}=X^iP^j-P^jX^i$, which generate the $p$-dimensional Euclidean subgroup $E(p)$ of isometries of $\Sigma_0$. The remaining $p+1$ generators, the (instant form) Hamiltonian $H=P^0$ and the boost generators $M^{0i}$, are thus dynamical. In the point-form of dynamics the kinematic subgroup is the Lorentz group generated by $M^{\mu\nu}$ while the four momenta $P^\mu$ are dynamical and complicated. Finally, in Dirac's front form of dynamics the kinematic subgroup is the group of transformations that leave the three-dimensional hyperplane $\Sigma_+$ invariant.
	\newpage

	\noindent We introduce light-like coordinates adapted to the front-form of dynamics,
	\be
	x^{\pm}=-x_{\mp}\equiv \frac{1}{\sqrt{2}}(x^0\pm x^p),
	\ee
	while we denote the transverse coordinates $\vec{x}_\perp$ by $x^a$, with $a=1,\cdots,p-1$. In such a frame, the non-zero components of the flat metric tensor are
	\be
	\eta_{ab}=\delta_{ab}\ \ ; \ \ \eta_{+-}=\eta_{-+}=-1.
	\ee
	The corresponding conjugate momenta take the form
	\be
	P^{\pm}=-P_{\mp}\equiv \frac{1}{\sqrt{2}}(P^0\pm P^p)=-i\partial^\pm=i\partial_\mp \ \ , \ \ P^a=-i\partial^a=-i\partial_a
	\ee
	and the canonical commutation relations read
	\be
	[X^\pm,P^\mp]=-i \ \ , \ \ [X^a,P^b]=i\delta^{ab}. 
	\ee
	
				\begin{figure}[!h]
					\centering
					\includegraphics[width=0.95\textwidth]{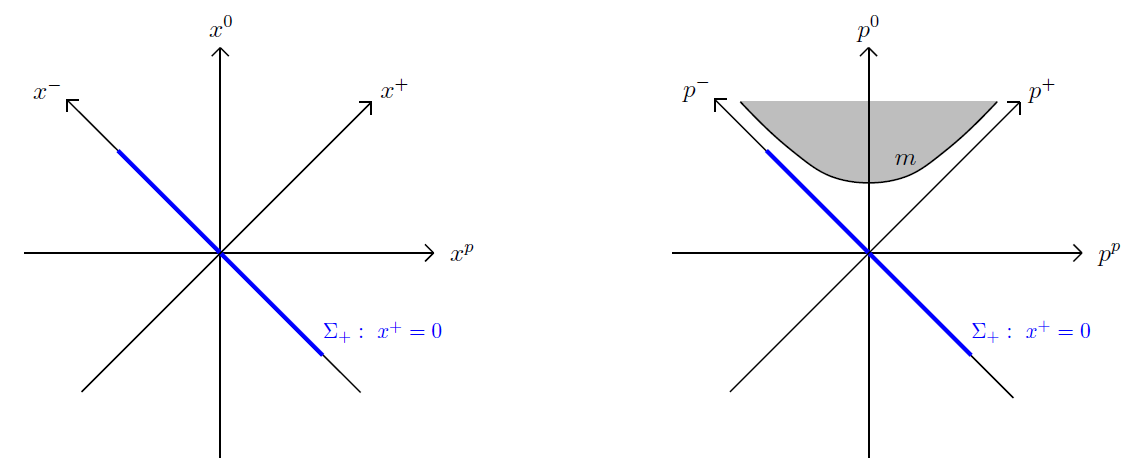}
					\caption{2D representation of the light-front coordinates, the initial hypersurface $\Sigma_+$ and the spectral condition for a single mass system.}
				\end{figure}
	
	It can be shown that the $\frac{1}{2}p(p+1)+1$ generators $P^+,P^a, M^{ab}, M^{+-}\equiv X^+P^--X^-P^+$ and $M^{+a}\equiv X^+P^a-X^aP^+$ realize transformations that leave $\Sigma_+$ invariant, so they are the kinematical generators in the front form of dynamics and will look simple. The remaining ones, the $p$ generators $P^-, M^{-a}\equiv X^-P^a-X^aP^-$, will be complicated in general and are the dynamical generators or Hamiltonians. For the sake of clarity, it is also good to express the transformed Lorentz generators in terms of the usual angular momenta $M^{ij}$ and boost generators $M^{0i}$. They read
	\be
	M^{+a}=\frac{1}{\sqrt{2}}(M^{0a}+M^{pa}) \ \ \ ,\ \ \ M^{-a}=\frac{1}{\sqrt{2}}(M^{0a}-M^{pa}) \ \ \ , \ \ \ M^{+-}=-M^{0p}.
	\ee
	Is is worth noticing that the front form of dynamics has the maximal number of kinematical generators, with one dynamical generator less compared to the instant or point forms, and therefore offers a priori a form of dynamics better suited for dealing with interactions in a relativistic setup.\\
	\indent   One of the most peculiar and powerful features of the light-front formulation is its manifest non-relativistic Galilean invariance, that is, the appearance of a Galilean subgroup once we express the Poincar\'e algebra in light-like coordinates. In order to see this, let's focus on the commutation relations of the $\frac{1}{2}p(p+1)$ generators $P^-, P^a, M^{+a}$ and  $M^{ab}$: 
	\ba
	&&[P^a,P^b]=0 \ \ \ , \ \ \ 	[P^a,P^-]=0 \ \ \ , \ \ \ 	[M^{ab},P^c]=i(\delta^{ac}P^b-\delta^{bc}P^a) \hspace{1.5cm} \no
	\ &&[M^{+a},P^-]=-iP^a \ \ \ , \ \ \ [M^{+a},P^b]=-i\delta^{ab} P^+ \ \ \ , \ \ \ [M^{+a},M^{+b}]=0 \no
	\ &&[M^{ab},M^{+c}]=i(\delta^{ac}M^{+b}-\delta^{bc}M^{+a}) \ \ \ , \ \ \ [M^{ab},P^-]=0 \no
	\ &&[M^{ab},M^{cd}]=i(\delta^{ac}M^{bd}-\delta^{bc}M^{ad}+\delta^{bd}M^{ac}-\delta^{ad}M^{bc}) .
	\ea
	We easily recognize the algebra for the Galilean group in $p$-dimensions $G(p,1)$ with its central extension. $P^-$ plays the role of the non-relativistic Hamiltonian generating (Galilean) time translations, 	$P^a$ are the $p$ generators of spatial translations, $M^{ab}$ are the usual $SO(p)$ generators of spatial rotations and $M^{+a}$ would be the $p$ generators of (Galilean) boosts. At the same time, $P^+$ 	is the central extension and may be seen as the non-relativistic mass.\\
	\indent This change of variables can be applied straightforwardly to conformal field theories (CFT). In this second case it is easy to see that the dilatation operator $D$ and the $K^+, K^a$ components of the special conformal transformations enlarge the set of kinematical generators, while $K^-$ will be an extra dynamical generator for a CFT quantized in the light-front. Equivalently, starting with the conformal algebra we will end up with a Schr\"odinger subgroup, that is, the symmetry group of the free Schr\"odinger equation. In addition to the previous Galilean generators we may consider two more generators, $\tilde{D}$ and $\tilde{K}$, defined by
	\be
	\tilde{D}=D-M^{+-} \ \ \ \mbox{and} \ \ \ \tilde{K}=\frac{1}{2}K^+=\frac{1}{2\sqrt{2}}(K^0+K^p),
	\ee
	and satisfying the following commutation relations:
	\ba
	&&[\tilde{D},P^a]=iP^a \ \ \ , \ \ \ 	[\tilde{D},P^-]=2iP^- \ \ \ , \ \ \ 	[\tilde{D},M^{ab}]=0 \ \ \ , \ \ \ 	[\tilde{D},M^{+a}]=-iM^{+a} \hspace{0.4cm}\no
	&&[\tilde{K},P^a]=iM^{+a} \ \ \ , \ \ \ 	[\tilde{K},P^-]=-i\tilde{D} \ \ \ , \ \ \ 	[\tilde{K},M^{ab}]=0 \ \ \ , \ \ \ 	[\tilde{K},M^{+a}]=0 \no
	&&[\tilde{D},\tilde{K}]=-2i\tilde{K} .
	\ea
	$\tilde{D}$ is a dilatation operator which, unlike the relativistic case, scales time and space differently. As a differential operator it reads $\tilde{D}=-(2t\partial_t+x^a\partial_a)$ so that
	\be
	x^a \to \lambda x^a \ , \ \ \ t \to  \lambda^2 t.
	\ee
	Similarly, $\tilde{K}$ acts something like the time component of the spatial conformal transformations for a relativistic CFT. It takes the form  $\tilde{K}=-(tx^a \partial_a+t^2\partial_t)$ and generates the finite transformations (parametrized by the scale $\lambda$)
	\be
	x^a \to \frac{x^a}{1+\lambda t} \ , \ \ \ t \to  \frac{t}{1+\lambda t} .
	\ee
	\noindent Up to this point we merely performed a change of coordinates, from cartesian to light-like, but the key difference between the various forms of dynamics comes from where do we decide to impose our initial conditions. In complete analogy with the situation in the instant form case, for fixed $x^+=0$ the generator $P^-=-i\partial^-=i\partial_+$ is no longer well-defined and must be expressed in terms of the remaining dynamical variables. Using Dirac's method of Lagrange multipliers and imposing the mass-shell conditions $P^\mu P_\mu=P^aP_a-2P^-P^+=-m^2$, we end up with
	\be
	P^-=\frac{P^aP_a+m^2}{2P^+}.
	\ee
	We can see at first sight that the light-front dispersion relation has several remarkable properties. As already pointed out by Dirac, this Hamiltonian does not contain a square root and the dependence on the transverse momenta $P^a$ is of the same quadratic form as in non-relativistic dynamics.	However, there is also a manifest singularity at $P^+ = 0$, which therefore is a somewhat peculiar point which needs extra care. Far from pretending to sweep this under the carpet, in what follows we will be concerned with this subtle issue and at the end of the day we will see how zero modes, that is, modes with zero longitudinal momentum $P^+$, play a prominent role in LFQ.
	
		\subsection{Light-front quantization of fields}
	Similar to the situation in equal-time quantization, the dynamical operators are calculated by integrating densities over the initial value surface. In particular, we may derive a representation of the Poincar\'e generators in terms of the initial surface and the canonical energy-momentum tensor
	\be
	T^{\mu\nu}=\frac{\partial \mathcal{L}}{\partial(\partial_\mu \Phi)}\partial^\nu \Phi-g^{\mu\nu} \mathcal{L},
	\ee
	with $\mathcal{L}$ being the Lagrangian density which depends on fields that are collectively denoted by $\Phi$ as usual. In particular, the Lorentz generators $P^\mu$ and $M^{\mu\nu}$ can be obtained	by integrating over either the space-like surface $\Sigma_0$,
	\ba \label{2generatorsIF}
	P^\mu&=&\int_{\Sigma_0}d^3x T^{0\mu} \ , \no
	M^{\mu\nu}&=&\int_{\Sigma_0}d^3x (x^\mu T^{0\nu}-x^\nu T^{0\mu}),
	\ea	
	or the light-front surface $\Sigma_+$,
	\ba \label{2generatorsFF}
	P^\mu&=&\int_{\Sigma_+}dx^-d\vec{x}_\perp T^{+\mu} \ , \no
	M^{\mu\nu}&=&\int_{\Sigma_+}dx^-d\vec{x}_\perp (x^\mu T^{+\nu}-x^\nu T^{+\mu}).
	\ea
	In both cases, it is easy to show that they generate the correct commutation relations with respect to the fields $\Phi$	and among themselves  \cite{Chang:1972xt}.
	\newpage
	
	\noindent In order to actually quantize the fields on the hypersurface $\Sigma_+$, we start by defining a four-momentum density canonically conjugate to $\Phi$
	\be
	\Pi^\mu=\frac{\partial \mathcal{L}}{\partial(\partial_\mu \Phi)}.
	\ee
	Using then the normal $n^\mu\propto\partial^-$ of $\Sigma_+$, we define the canonical momentum density 
	\be
	\pi\equiv n \cdot \Pi=\partial^+\Phi=-\partial_-\Phi,
	\ee 
	which, compared to equal-time quantization, is very peculiar concerning that it doesn't involve a light-front time derivative $\partial_+$. Therefore, $\pi$ is a dependent quantity which does not provide any additional information as an initial condition on $\Sigma_+$. Obviously, the reason is that the normal $n^\mu$, being light-like, is both normal and tangent to $\Sigma_+$. At the end of the day, it can be shown that the equal-(light-front)time canonical commutation relations are \cite{Chang:1972xt, Heinzl:1998kz}
	\be
	[\Phi(x),\pi(y)]_{x^+=y^+=0}=i\delta(x^--y^-)\delta^{p-1}(\vec{x}_\perp-\vec{y}_\perp).
	\ee 
	As the independent quantities are the fields themselves, we may invert the derivative $\partial^+$ and obtain the fundamental commutation relations that we have to impose on $\Sigma_+$,
	\be
	[\Phi(x),\Phi(y)]_{x^+=y^+=0}=-\frac{i}{2}\mbox{sgn}(x^--y^-)\delta^{p-1}(\vec{x}_\perp-\vec{y}_\perp).
	\ee
	
	\subsection{The light-front vacuum and the zero mode problem}
	The spectral condition (see Figure 1) of relativistic quantum field theory assumes that the spectrum of the four-momentum operator is contained within the closure of the forward light-cone. That is, the $d$-momentum of any physical observable particle obeys
	\be \label{W0}
	P^2\leq 0 \ , \ \ P^0\geq 0.
	\ee
	From the above condition we infer directly that
	\be
	P_0^2-P_p^2\geq \vec{P}^2_\perp\geq 0 \ \ \ \ \mbox{or equivalently} \ \ \ \ P^0\geq |P^p|.
	\ee
	This implies that the longitudinal light-front momentum $P^+$ satisfies
	\be
	P^+=\frac{1}{\sqrt{2}}(P^0+P^p)\geq \frac{1}{\sqrt{2}}(|P^p|+P^p)\geq 0,
	\ee
	and we thus arrive to a very important kinematical constraint characteristic of the front form of dynamics: physical states must have non-negative longitudinal light-front momentum,
	\be
	\langle \mbox{phys}|P^+|\mbox{phys}\rangle \geq 0.
	\ee
	Mostly all the distinctive features of LFQ stem directly from the fact that $P^+$ is a kinematical generator and at the same time a positive semi-definite operator with its spectrum bounded from below. In addition, Poincar\'e/conformal invariance dictates that the light-front vacuum $|0\rangle$ must be annihilated by the kinematic generators of the Poincar\'e/conformal group and, in particular,
	\be \label{2LFvacuum}
	P^+|0\rangle=0.
	\ee
	The LFV is thus an eigenstate of $P^+$ with the lowest possible eigenvalue. It is in this sense that is is usually said that the LFV is trivial, since we can always specify the eigenvalues of $P^+$ without solving the dynamics and thus the vacuum can be defined kinematically. Upon such a trivial vacuum we can construct the physical Fock space in terms of fields whose Fourier components with $p^+>0$ and $p^+<0$ must be interpreted as creation and annihilation operators, respectively. For instance, this can be seen easily by analyzing the $x^+$ representation of a scalar field propagator \cite{Chang:1968bh}
	\ba \label{2propagator}
	&\Delta_F(x^+,p^+,\vec{p_\perp})&=\frac{-1}{|2p^+|}\theta(x^+p^+)\exp\Big[-ix^+\frac{\vec{p}^{\ 2}_\perp+m^2}{2p^+}\Big] \ \mbox{ if } \ p^+\neq 0\no
		&\Delta_F(x^+,p^+,\vec{p_\perp})&=\frac{1}{\vec{p}^{\ 2}_\perp+m^2}\delta(x^+) \ \mbox{ if } \ p^+= 0.
	\ea
	Apart from the case  $p^+=0$, which will be discussed later, we see how for $p^+>0$,  $\Delta_F(x^+,p^+,\vec{p_\perp})$ is nonzero only when $x^+>0$, and for $p^+<0$, only when $x^+<0$. Interpreting $\frac{\vec{p}^{\ 2}_\perp+m^2}{2p^+}$ as the single-particle energy, then (\ref{2propagator}) tells us  that positive-energy states propagate forward in $x^+$ and negative-energy states propagate backward. As usual, the latter may also be regarded as antiparticles propagating forward in $x^+$. It is worth noticing  that although $p^-$ is the conjugate to $x^+$, $\theta(x^+p^+)$ appears in (\ref{2propagator}) instead of $\theta(x^+p^-)$, in sharp contrast to $\theta(x^0p^0)$ in equal-time quantization.\\
	\indent With an eye on phenomena such are vacuum polarization or spontaneous symmetry breaking, it is of concern wether there is or not vacuum degeneracy. 	In order to show this, let's proceed by reduction to absurdity and assume there is another state, $|p^+ = 0,\alpha\rangle$, having the same eigenvalue $p^+ = 0$ as the vacuum and denoted by an extra label $\alpha$. If so, it must be possible to create such a state from the vacuum acting with some operator $\mathcal{O}_\alpha$,
	\be \label{2degeneracy}
	|p^+ = 0,\alpha\rangle=\mathcal{O}_\alpha|0\rangle,
	\ee
	where $\mathcal{O}_\alpha$ must not produce any longitudinal light-front momentum. Now the situation is completely different from the usual one in equal time quantization, where typically one can always act with a given combination of Fock operators $a^\dagger(\vec{k})$ and $a^\dagger(-\vec{k})$ such that contributions from modes
	with positive and negative momenta cancel each other. The eigenvalues of $P^+$ being non-negative, here the problem boils down to the question whether there are Fock operators carrying null light-front momentum, i.e. $a^\dagger(k^+ = 0)$. For the moment we deliberately disregard the \emph{zero mode} with $p^+=0$, which will play a central role and is to be discussed later. Accordingly, the only remaining possibility is that, if $\mathcal{O}_\alpha$ contains any creation operator $a^\dagger(k^+>0)$ carrying non-vanishing longitudinal momentum, there must be the appropriate number of annihilation operators annihilating the same amount of momentum. Thus, after Wick ordering, any operator $\mathcal{O}_\alpha$ satisfying (\ref{2degeneracy}) must have the general form
	\ba \label{2generalform}
	\mathcal{O}_\alpha=\langle 0|\mathcal{O}_\alpha|0\rangle&+&\int\limits_{k^+>0}dk^+f_1(k^+)a^\dagger(k^+)a(k^+)\no
	&+&  \int\limits_{p^+>0}dp^+\int\limits_{k^+>0}dk^+f_2(k^+,p^+)a^\dagger(p^++k^+)a(p^+)a(k^+)\no
	&+&  \int\limits_{p^+>0}dp^+\int\limits_{k^+>0}dk^+\tilde{f}_2(k^+,p^+)a^\dagger(p^+)a^\dagger(k^+)a(p^++k^+) + \cdots
	\ea
	where we omitted for simplicity the contributions from Fock operators carrying transverse momenta $\vec{k}_\perp$, which are a priori unconstrained and analogous to the usual equal-time quantization. It follows then that the LFV is an eigenstate of $\mathcal{O}_\alpha$,
	\be 
	\mathcal{O}_\alpha|0\rangle=\langle 0|\mathcal{O}_\alpha|0\rangle|0\rangle,
	\ee
	and thus we conclude that the light-front vacuum defined by (\ref{2LFvacuum}) is unique and not degenerate (if we ignore the zero modes). As a matter of fact, any quantity obtained by integrating some functional of the fields over configuration space,
	\be
	F[\Phi]=\int dx^-d^2\vec{x}_\perp\mathcal{F}[\Phi]
	\ee
	will be of the general form (\ref{2generalform}), since integration in configuration space produces essentially Dirac delta functions and impose momentum conservation:
	\ba
	\int \frac{dx^-}{2\pi}e^{-ix^-(\sum_jp^+_j)}&=&\delta\Big(\sum\limits_jp_j^+\Big) \ , \ \ \mbox{and} \no
	\int \frac{d\vec{x}_\perp}{(2\pi)^{p-1}}e^{i\vec{x}_\perp(\sum_j\vec{p}_{\perp j})}&=&\delta^{(p-1)}\Big(\sum\limits_j\vec{p}_{\perp j}\Big).
	\ea
	Of course, the most important examples for such quantities are the Lorentz generators, as it is clear from their representations in terms of the energy-momentum tensor (\ref{2generatorsFF}). This implies in particular that the LFV is also an eigenstate of the fully interacting light-front Hamiltonian,
	\be \label{2LFVeigenstateH}
	P^-|0\rangle=\langle0|P^-|0\rangle|0\rangle.
	\ee
	Alternatively, this can also be seen directly from the algebra, since
	\be
	P^+P^-|0\rangle=P^-P^+|0\rangle=0
	\ee
	tells us that $P^-|0\rangle$ is a state with $p^+ = 0$, that is, proportional to the LFV. The particular value of $\langle0|P^-|0\rangle$ may seem unimportant at first, as it only defines the zero of light-front energy, but it will turn out to be crucial in our forthcoming conjecture. On general grounds, a constant non-vanishing (light-front) vacuum expectation value of the (light-front) Hamiltonian $P^-$ will break spontaneously some of the supersymmetries and those bosonic symmetries whose generators don't commute with $P^-$.	In summary, if we neglect the contribution of the zero modes, the light-front vacuum is non-degenerate and trivial, in the sense that there is no vacuum but the Fock vacuum defined by the truly absence on any particle. It is then worth noticing that this suppose a big difference compared to equal time quantization, where the Fock vacuum of an interacting field theory is not an eigenstate of the full Hamiltonian $H=P^0$ and we say that the (instant-form) vacuum is nontrivial. \\
	\indent Such a trivial vacuum, on the other hand, confronts  the usual picture of complicated non-perturbative vacuum structure in the equal-time quantization corresponding to confinement,  spontaneous symmetry breaking, etc... Since the vacuum is proved trivial and non-degenerate, the only possibility to realize such phenomena would be through the complicated structure of the zero modes, which we have ignored so far. If there exists a zero mode state degenerate with the LFV (\ref{2LFvacuum}), then the trivial Fock space vacuum may not be the true vacuum. In order to appreciate the effect of the zero modes we follow \cite{Chang:1968bh} and consider the diagrams presented in Figure 2 for the case of a $\phi^3$ model. The fact that $p^+$ is conserved at each vertex and that a line with $p^+>0$ ($<0$) must point forward (backward) in $x^+$ impose limits for the $p'^+$ integration of 2(a) (in particular, $p'^+\in [-p^+,0]$) and also tells us that the diagram of 2(b) is forbidden. Similarly, if we were to ignore the contribution of the zero modes, we would conclude (erroneously) that all vacuum diagrams should vanish as well\footnote{In the seminal work of Weinberg \cite{Weinberg:1966jm}, the point $p^+=0$ was always ignored, and the contribution from vacuum diagrams was thus lost.}. Let us analyze the lowest-order vacuum diagram of Fig. 2(c) and see why the point $p^+=0$ may not be ignored.

		\begin{figure}[!h]
			\centering
			\includegraphics[width=0.80\textwidth]{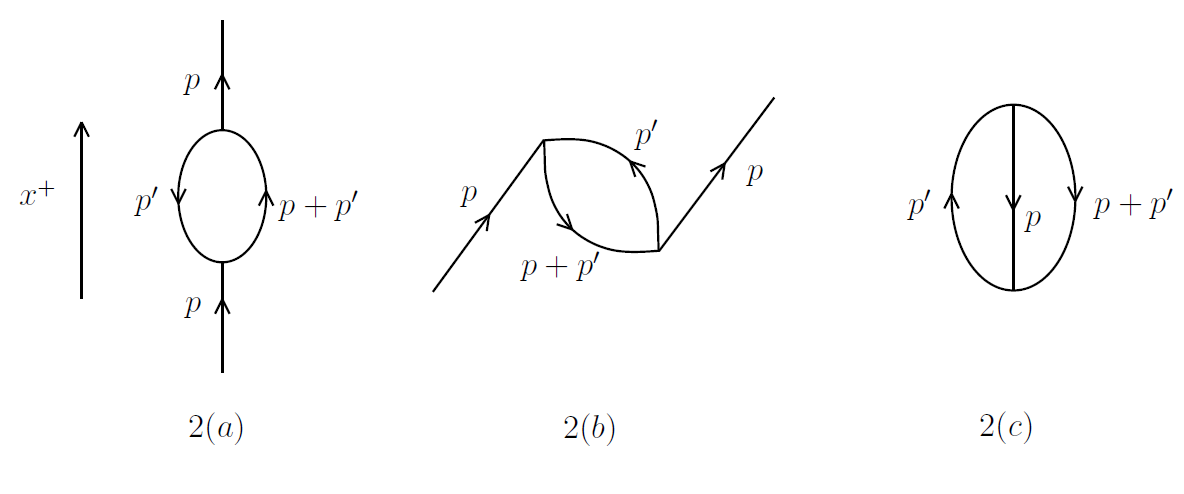}
			\caption{2(a) and 2(b) depict two second-order diagrams for the self-energy with two different $x^+$ orderings. 2(c) is the lowest-order vacuum bubble diagram.}
		\end{figure}
		
Figure 2(c) may be obtained from figure 2(a) by closing the two external lines and integrating over $p$. This vacuum diagram is then expressed as
\be
E=\int d^{d-2}\vec{p}_\perp dp^+dp^-\frac{\Sigma(p^2)}{-2p^+p^-+\vec{p}^{\ 2}_\perp+m^2-i\epsilon},
\ee
where the self-energy $\Sigma(p^2)$ is given by
\be
\Sigma(p^2)=\frac{i}{2}(ig)^2(2\pi)^{-d}\int d^{d-2}\vec{p'}_\perp dp'^+dp'^- \Delta_F(p)\Delta_F(p+p').
\ee
In order to show manifestly the contribution of the zero modes it proves useful to write $\Sigma(p^2)$ in the form of a Fourier transform
\be
\Sigma(p^+p^-)=\int d\lambda F(\lambda) e^{i\lambda p^+p^-}.
\ee
After a little algebra and using several times the identity
\be
(x+i\epsilon)^{-1}=-i\int_0^\infty d\xi e^{i\xi(x+i\epsilon)},
\ee
$F(\lambda)$ is found to be
\be
F(\lambda)=\frac{i}{2}(ig)^2(2\pi)^{1-d}\int d\xi_1d\xi_2 \delta(\lambda(\xi_1+\xi_2)-\xi_1\xi_2)e^{-im^2\xi_1\xi_2/\lambda}
\ee
and
\be \label{2LFVenergy1}
E=-2\pi i\int dp^+\Big[\int_0^\infty  d\xi d\lambda F(\lambda) (\lambda+\xi)^{-1} e^{-im^2\xi} \Big]\delta(p^+).
\ee
Similar computations can be straightforwardly generalized to other theories and to other kinds of matter with similar results. In conclusion, at the perturbative level zero modes may be ignored unless one deals with a vacuum diagram, which is in general nonzero due to the factor $\delta(p^+)$. Zero modes are thus the only responsables for the non-vanishing energy density of the light-front vacuum, $\mathcal{P}^-=\langle T^{--}\rangle$. Thus, when we write the light-front Hamiltonian in terms of the energy-momentum tensor as
	\be \label{2dynamicalmatter}
	P^-=\int_{\Sigma_+}dx^-d\vec{x}_\perp T^{+-},
	\ee
we need to take in consideration that we are integrating only the contribution from the dynamical matter, that is, those degrees of freedom corresponding to matter that crosses the initial surface $\Sigma_+$ once. On the other hand, zero modes with $p^+=0$ don't ever cross $\Sigma_+$ and their net effect translates, among others, into a non-vanishing vev for $P^-$ of the form
\be \label{2zeromodes}
\langle P^- \rangle =\int_{\Sigma_-}dx^+d\vec{x}_\perp \langle T^{--}\rangle,
\ee
that we have to add into (\ref{2dynamicalmatter}). This is coherent with the known fact that, most generally, the zero mode operator is not an independent degree of freedom but obeys a constraint equation. In this sense, the zero mode needs to be seen as a modification of the Hamiltonian rather than as a dynamical field. Of course, this changes completely the paradigm for (instant form) vacuum effects like spontaneous symmetry breaking or confinement. Multiple vacua are now replaced by multiple Hamiltonians and picking a given Hamiltonian defines the theory and phase in the same sense that picking a particular vacuum state defines the theory and phase in equal-time quantization\footnote{More information on the zero mode issue may be found in \cite{Yamawaki:1998cy}, chapter 7  of \cite{Brodsky:1997de} and chapter 7 of  \cite{Heinzl:1998kz}.}. 

	\subsection{The infinite-momentum frame}	
	Dirac's forms of relativistic dynamics remained almost completely forgotten during all the fifties and the sixties, but similar constructions were re-invented several times. Specially interesting for us will be the so-called \emph{infinite-momentum frame} approach (IMF), first appeared in the work of Fubini and Furlan \cite{Fubini:1964boa} in an attempt to simplify as much as possible certain sum rules involved in the renormalization of the coupling constants of generic QFTs. Soon after that Weinberg \cite{Weinberg:1966jm} realized that applying ``old-fashioned'' Hamiltonian perturbation theory in a reference frame with infinite total momentum $P$ exhibits many desirable properties. Most notably, if we let $P\to\infty$, every individual diagram either approaches a finite limit or vanishes, the vanishing diagrams being just the ones, like vacuum fluctuations, that caused the worst trouble (while neglecting the contribution form zero modes). Later Susskind \cite{Susskind:1967zza,Susskind:1967rg} pointed out new simplifications characteristic of the infinite-momentum limit and draw attention to the fact that this limiting form of the theory possess Galilean invariance with respect to motions transverse to the direction in which the momentum is infinite. Other relevant contributions where the works of Kogut and Soper \cite{Kogut:1969xa} and of Chang and Ma \cite{Chang:1968bh}, where it was partially proved, for a $\phi^3$-theory and for quantum electrodynamics, that LFQ was in fact equivalent to taking the infinite-momentum limit while avoiding the limiting procedures characteristic of the IMF approach. All in all, the infinite momentum frame may be useful as an intuitive tool, but the limit $P\to \infty$ is not a rigorous limit in general, and the need to boost the instant-form wave function introduces complexities. In retrospect, it is much more rigorous to work directly with the rules of LFQ while using some of the ideas of the IMF for clarifying the picture and for heuristic reasoning. On these lines, part of our initial motivation was precisely to translate the results of \cite{Cvetic:1998jf,Brecher:2000pa} into a LFQ language.
	
	\section{AdS with a propagating pp-wave as the gravitational dual of the light-front vacuum state}	
	We are now ready for addressing the main questions posed in the introduction: what is the geometry dual to the light-front vacuum state (\ref{2LFvacuum}) of a given SCFT in the limit of strong coupling and large central charges? How do we implement holographically the effect (\ref{2zeromodes}) of  the zero modes?
	\subsection{Brane-wave spacetimes and their near-horizon limit}
	In \cite{Cvetic:1998jf,Brecher:2000pa} spacetimes describing pp-waves propagating on non-dilatonic branes tangentially to the worldvolume were studied in detail. On the one hand, it was observed that for non-extremal configurations of this kind the effect of the inclusion of the pp-wave is globally equivalent to performing a Lorentz boost along the direction of propagation of the wave if such direction is uncompactified, while if the direction is wrapped on a circle the equivalence is only valid locally. For the case of BPS saturated $p$-branes, on the other hand, the inclusion of the pp-wave leads to a metric that is not even locally equivalent to the one where there is no wave, preserving only $1/4$ of the supersymmetry and with a singularity at the horizon of the brane. The reason behind this distinction lies in the fact that in the BPS limit the boost that relates the two metrics (those with and without a wave) becomes singular, corresponding to a boost at the speed of light. Taking the decoupling limit of these extremal brane-wave spacetimes gives a pp-wave propagating in AdS, the simplest example being the Kaigorodov spacetime \cite{Kaigorodov}. Finally, and what is most relevant for our interests, it was conjectured that supergravity in Kaigorodov spacetime is dual to a CFT in the infinite momentum frame with constant momentum density and in the large $N$ and strong coupling limit.\\
	\indent Let's now revisit the main aspects of their derivation. We start with an ansatz for the metric of a non-dilatonic extremal $p$-brane with a pp-wave propagating along a worldvolume direction, which we take to be the negative $x^p$-direction\footnote{Heuristically, the idea behind this choice is that light-front quantization imposing initial conditions on $\Sigma_+$ can be understood as the kind of physics that an observer moving at the speed of light in the negative $x^p$-direction would experience. Accordingly, such an observer would see everything moving at the speed of light in the positive $x^p$-direction, but for the particular case of the LFV state defined by (\ref{2LFvacuum}), there is no dynamical matter propagating in the positive $x^p$-direction and there are only zero modes propagating in the negative $x^p$-direction. Thus, the pp-wave wave will be somehow  the holographic dual to the zero mode sector in the CFT.}. Such a solution will be of the general form:
	\be \label{3ansatz}
	ds^2=H(r)^{-2/d}(-2dx^+dx^-+F(x^+,x^a,r)dx^{+2}+dx^a dx^a)+H(r)^{2/\tilde{d}}(dr^2+r^2d\Omega^2_{\tilde{d}+1}),
	\ee
	where $d=p+1$, $\tilde{d}+2=D-d$ \ is the codimension of the brane, $D$ is the number of spacetime dimensions of the bulk and
	\be
	\frac{2}{d}+\frac{2}{\tilde{d}}=1
	\ee
	for the branes we are considering\footnote{For simplicity, we focus on non-dilatonic branes only, that is, we concentrate on the extremal D1-D5 system and D3-branes of type IIB theory and on the extremal M2- and M5-branes of M-theory.}. A priori the function $F(x^+,x^a,r)$ may depend on both worldvolume and transverse coordinates, but cannot depend on both $x^+$ and $x^-$ as it has to solve a wave equation. It can be verified that (\ref{3ansatz}) solve the Einstein equations provided that the function $H(r)$ is harmonic on the transverse space,
	\be
	H(r)=1+\frac{k}{r^{\tilde{d}}},
	\ee
	and the function $F(x^+,x^a,r)$ obeys
	\be \label{3Feq}
	\frac{\partial^2 F}{\partial r^2}+\frac{(\tilde{d}+1)}{r}\frac{\partial F}{\partial r}+H(r)\nabla^2_{\perp}F=0,
	\ee
	where $\nabla^2_{\perp}$ stands for the Laplacian with respect to the flat transverse coordinates $x^a$. If we focus now on the near horizon limit of (\ref{3ansatz}), the $1$ in the harmonic function $H$ can be dropped as usual. Defining the holographic coordinate $z$ as
	\be
	z=\frac{d}{\tilde{d}}k^{1/2}r^{1-\tilde{d}/2},
	\ee
	the near horizon metric takes the form
	\be \label{3NHansatz}
	ds^2=\frac{L^2}{z^2}\Big(-2 dx^+dx^-+F(x^+,x^a,z)dx^{+2}+dx^a dx^a+dz^2\Big)+(\tilde{d}/d)^2L^2d\Omega^2_{\tilde{d}+1},
	\ee
	where $L=(\tilde{d}/d)k^{1/\tilde{d}}$.	In the near horizon limit and in terms of the new coordinate, the equation satisfied by $F$ is now
	\be \label{3NHF}
	z^{d-1}\frac{\partial}{\partial z}\Big(\frac{1}{z^{d-1}}\frac{\partial F}{\partial z}\Big)+\nabla^2_{\perp}F=0.
	\ee
	With these rearrangements we recognize now the metric describing a pp-wave traveling along $AdS_{d+1}$ and characterized by the function $F(x^+,x^a,z)$. It is worth noticing that the conformal boundary at $z=0$ is no longer conformal to Minkowski space in general but otherwise depends on the asymptotic value $F(x^+,x^a,0)$. It was shown in great detail in \cite{Brecher:2000pa} that there are no non-trivial and non-singular metrics of the form (\ref{3ansatz}). Focusing only in the near-horizon metric, there is indeed a particular solution of (\ref{3NHansatz}, \ref{3NHF}) free of singularities, but further inspection shows it just describes AdS in unusual coordinates and thus such a solution is pure gauge. Considering the whole brane-wave metric or its near horizon limit, it was shown to be no non-trivial and non-singular solutions of either (\ref{3ansatz}, \ref{3Feq}) or (\ref{3NHansatz}, \ref{3NHF}).
	\subsection{Energy-momentum tensor of the dual field theory}
	We have shown how the near-horizon geometry of a pp-wave propagating tangential to a extremal non-dilatonic $p$-brane translates into a pp-wave propagating in AdS and we will now consider how the AdS/CFT correspondence works for such cases. It is worth mentioning that if such a duality applies, the CFT would provide a resolution for the singularity at the horizon. In order to go directly to the point, we will focus by now on the expectation value of the CFT energy-momentum tensor and will leave the matching of symmetries and other checks for later.\\
	\indent By means of the standard boundary counterterm method \cite{Henningson:1998gx, Balasubramanian:1999re, Emparan:1999pm, Kraus:1999di} and using the near-horizon metric (\ref{3NHansatz}), we can compute the expectation value of the energy-momentum tensor at the conformal boundary. We then perform a Weyl transformation in order to put the boundary metric as
	\be
	ds^2=-2dx^+dx^-+F(x^+,x^a,0)dx^{+2}+dx^a dx^a,
	\ee
	which is manifestly flat only if $F(x^+,x^a,0)=0$. In the new conformal frame, the only non-vanishing component of the CFT energy-momentum tensor turns out to be
	\be
	\langle T^{--}\rangle=	\langle T_{++}\rangle=\frac{L^{d-1}}{16\pi G_{d+1}}\lim\limits_{z\to 0}\frac{1}{z^{d-1}}\Big[\frac{\partial F}{\partial z}-\frac{z}{d-2}\nabla^2_{\perp}F-\frac{z^3}{(d-2)^2(d-4)}(\nabla^2_{\perp})^2F+\cdots\Big],
	\ee
	where we have included the first three boundary counterterms for arbitrary $d$. As usual, it is much more illuminating to write everything in terms of the parameters of the dual CFT. For the cases we are considering, we get
	\ba
	\langle T^{--}\rangle&=&\frac{c}{24 \pi}\lim\limits_{z \to 0}\frac{1}{z}\frac{\partial F}{\partial z}, \no
	\langle T^{--}\rangle&=&\frac{\sqrt{2}N^{3/2}}{24 \pi}\lim\limits_{z \to 0}\frac{1}{z^2}\Big[\frac{\partial F}{\partial z}-z\nabla^2_{\perp}F\Big], \no
	\langle T^{--}\rangle&=&\frac{N^{2}}{8 \pi^2}\lim\limits_{z \to 0}\frac{1}{z^3}\Big[\frac{\partial F}{\partial z}-\frac{z}{2}\nabla^2_{\perp}F\Big], \no
	\langle T^{--}\rangle&=&\frac{N^{3}}{3 \pi^3}\lim\limits_{z \to 0}\frac{1}{z^5}\Big[\frac{\partial F}{\partial z}-\frac{z}{4}\nabla^2_{\perp}F-\frac{z^3}{32}(\nabla^2_{\perp})^2F\Big]
	\ea
	for $d=2,3,4$ and $6$ respectively. $c$	denotes de central charge of the 2D CFT while $N$ is the number of branes present before the decoupling and near horizon limits, which in turn corresponds to the rank of the gauge group of the dual field theory for $d>2$. It is immediate to see that both the induced metric at the conformal boundary and the expectation value of the energy-momentum tensor depend on the particular form of the function $F$ that characterizes the pp-wave.\\
	\indent Solutions of  (\ref{3NHansatz}) represent gravitational waves propagating in surfaces with constant holographic coordinate $z$ at the speed of light in the negative $x^p$ direction and with an amplitude depending upon $z$ and $\vec{x}_\perp$. Fourier analyzing (\ref{3NHF}) in the transverse coordinates gives $F\propto e^{i\vec{k}_\perp \cdot \vec{x}_\perp}$, where $\vec{k}_\perp$ could in principle depend on $x^+$. If $k_\perp$ is real, solutions would propagate faster than light for a given $z$ and may be seen as tachyons, which in turn denote an instabilty of the system. Taking $\vec{k}_\perp^2=-m^2$, we are led to the equation
	\be \label{3Fourier}
	z^{d-1}\frac{\partial}{\partial z}\Big(\frac{1}{z^{d-1}}\frac{\partial F}{\partial z}\Big)+m^2F=0,
	\ee
	which is immediate for $m^2=0$ and can be solved with Bessel functions for $m^2\neq 0$ 	(see \cite{Chamblin:1999cj} for more details). Taking the $z\to 0$ limit, the two independent solutions of (\ref{3Fourier}) scale either like $z^d$ or like $z^0$. As usual \cite{Balasubramanian:1998sn, Balasubramanian:1998de}, solutions proportional to $z^d$ give a finite expectation value, come from normalizable bulk modes and correspond to finite deformations of the state of the dual CFT. On the other hand, solutions proportional to $z^0$ at the boundary give in general a divergent expectation value, are associated with non-normalizable bulk modes and correspond to a change in the parameters of the UV Lagrangian of the CFT, in this case a modification of the boundary metric. In the present work we are interested only in solutions proportional to $z^d$ at the boundary, for which the conformal boundary metric is just Minkwonski space and the CFT is in a quantum state with non-vanishing $\langle T^{--}\rangle$, i.e. there is a finite null momentum density in the dual field theory. For clarity reasons, we may concentrate on the simplest solution
	\be
	F(x^+,x^a,z)=F(z)=\mu^dz^d,
	\ee
	where $\mu$ is just a constant with dimensions of energy in natural units. It is the $AdS$ analogue of the simplest vacuum pp-wave,
	namely, the homogeneous pp-wave in flat space. For this simplified setup, the boundary energy-momentum tensor describes a disturbance of matter propagating at the speed of light in the negative $x^p$-direction with a fixed constant wave profile whose magnitude is controlled by the scale $\mu$. In turn, the constant null momentum density $\mathcal{P}^-=\langle T^{--}\rangle$ is given by
		\be \label{3nulldensity}
		\mathcal{P}^-=\frac{c\mu^2}{12 \pi} \ \ , \ \ \mathcal{P}^-=\frac{\sqrt{2}N^{3/2}\mu^3}{8 \pi} \ \ , \ \ \mathcal{P}^-=\frac{N^2\mu^4}{2 \pi^2} \ \ \mbox{and} \ \ 	\mathcal{P}^-=\frac{2N^3\mu^6}{\pi^3}
		\ee
	for $d=2,3,4$ and $6$ respectively.\\
	 \indent In four dimensions, this particular pp-wave propagating in AdS spacetime was first discussed by Kaigorodov in \cite{Kaigorodov} and hence is known in the literature as the (four-dimensional) Kaigorodov spacetime $K_4$. Its generalization to $D$-dimensions, denoted $K_D$, was presented in \cite{Cvetic:1998jf} and fully covered in their appendix A. Heuristically, $K_D$ can be regarded as a singular infinitely boosted version of $AdS_D$ such that the infinite boost in the bulk induces an infinite boost on the conformal boundary. Thus, with the preceding discussion, the authors of \cite{Cvetic:1998jf,Brecher:2000pa} conjectured that supergravity theory in the Kaigorodov spacetime (times a compact manifold) is dual to a strongly coupled large $N$ SCFT in an infinitely boosted frame and in a given state with constant momentum density. \\
	\indent Now, with everything said so far, we arrive at the present work's raison d'\^etre: what we claim is that the Kaigorodov spacetime should be better interpreted not as the gravity dual of a SCFT in an infinitely boosted frame and in a certain unspecified state but as as the dual of a field theory quantized in the light-front and in the light-front vacuum state. The non-vanishing null momentum density captured holographically  (\ref{3nulldensity}) ultimately leads to a non-vanishing vev for the light-front Hamiltonian $P^-$, which is perfectly coherent with the contribution (\ref{2LFVenergy1}-\ref{2zeromodes}) of the zero modes that we discussed in the previous section. This way, it results very appealing to interpret the function $F$ as the AdS/CFT implementation of the zero modes and (\ref{3NHF}) as a kind of strong-coupling zero mode constraint equation. Furthermore, we argue that it is more natural and rigorous to interpret the dual field theory as quantized in the light-front rather than in an infinite-momentum frame since in the gravity side we have not performed any $P \to \infty$ limit.\\
	\indent Although at a first glance this may seem as only paraphrasing the work of \cite{Cvetic:1998jf,Brecher:2000pa}, it have far reaching consequences. To start with, it is fair to say that light-front quantization is way more formal and better understood than the infinite-momentum frame approach. Furthermore, identifying (\ref{3NHF}) as a zero mode constraint equation will allow us further studies. In particular, it would be interesting to reformulate vacuum phenomena like spontaneous symmetry breaking holographically and in terms of zero modes. Finally, being more precise about which is the particular state in the dual field theory that corresponds to Kaigorodov spacetime allows us to perform further quantitative as well as qualitative checks for the conjecture. In the following section we present such a sanity check.
	
	\newpage
	\subsection{Matching of symmetries}
	A key and necessary ingredient for a holographic correspondence to hold is that the global unbroken symmetries of the two theories must be identical. The metric for the ($d+1$)-dimensional Kaigorodov spacetime $K_{d+1}$ can be written in Poincar\'e coordinates as
	\be \label{3Kaigorodov}
	ds^2=\frac{L^2}{z^2}(-2dx^+dx^-+\mu^d z^d dx^{+2}+dx^a dx^a+dz^2),
	\ee
	which shows clearly that Kaigorodov spacetime is asymptotic to $AdS_{d+1}$ near the boundary (as $z \to 0$) and exhibits a pp-wave curvature singularity near the Poincar\'e horizon ($z \to \infty$). It was shown in \cite{Chamblin:1999cj,Cvetic:1998jf,Brecher:2000pa} that this metric has the following $\frac{1}{2}(p+1)(p+2)+1$ Killing vector fields (where we have adapted their results to our notation):
	\ba \label{3killings}
	&&\mathcal{P}_+=\frac{\partial}{\partial x^+} \ , \ \ \ \mathcal{P}_-=\frac{\partial}{\partial x^-} \ , \ \ \ \mathcal{P}_a=\frac{\partial}{\partial x^a} \no
	&&\mathcal{M}^{ab}=x^a\frac{\partial}{\partial x^b}-x^b\frac{\partial}{\partial x^a} \no
	&&\mathcal{M}^{+a}=x^+\frac{\partial}{\partial x^a}+x^a\frac{\partial}{\partial x^-}\no
	&&\mathcal{J}=z\frac{\partial}{\partial z}+x^a\frac{\partial}{\partial x^a}+\frac{1}{2}(d+2)x^-\frac{\partial}{\partial x^-}-\frac{1}{2}(d-2)x^+\frac{\partial}{\partial x^+}=\no
	&&\hspace{0.48cm}=z\frac{\partial}{\partial z}+x^a\frac{\partial}{\partial x^a}+x^-\frac{\partial}{\partial x^-}+x^+\frac{\partial}{\partial x^+}+\frac{d}{2}\Big(x^-\frac{\partial}{\partial x^-}-x^+\frac{\partial}{\partial x^+}\Big).
	\ea
	Now we need to prove that the isometries of $K_{d+1}$ generated by these Killing vectors are in one to one correspondence with the bosonic symmetries of the dual field theory in the light-front vacuum state. On the one hand, the LFV is defined kinematically by demanding that all kinematic generators must have a vanishing vacuum expectation value. On the other hand, a non-vanishing vev of the light-front Hamiltonian (\ref{2zeromodes}) breaks spontaneously some of the symmetries. In particular, it is easy to see from the algebra that it breaks the symmetries generated by $M^{+-}, D, K^+$ and $K^a$ while preserving those generated by $P^+, P^a, M^{ab}, M^{+a}$ and the combination $J\equiv D+M^{+-}$. Thus, the bosonic symmetries in the field theory side are those symmetries that preserve the vanishing vevs
	\be \label{3vevs}
	\langle P^+ \rangle=	\langle P^a \rangle=	\langle M^{ab} \rangle=	\langle M^{+a} \rangle=	\langle J \rangle=0.
	\ee
	Again, it is immediate from the algebra that the symmetries preserving (\ref{3vevs}) are the ones generated by 
	$P^+, P^-, P^a, M^{ab},  M^{+a}$ and $J$, and it is clear that the Killing vector fields of $K_{d+1}$ (\ref{3killings}) induce precisely these symmetries on the boundary theory.\\
	 \indent Some of these symmetries have an obvious interpretation. The $P^\mu$ and $M^{ab}$ generators just give boundary translations and transverse rotations, while $M^{+a}=\frac{1}{\sqrt{2}}(M^{0a}+M^{pa})$ is a combination of a boost in the $x^a$-direction, a rotation in the $x^ax^p$-plane and a boost in the $x^p$-direction. The remaining generator $J=D+M^{+-}$ involves an unbroken conformal symmetry and its implementation needs some clarification. It can be shown that the action of the symmetry generated by $J$ is
	 	\be
	  x^a\to {x^a}'=\lambda x^a \ , \ \ x^+\to {x^+}'=\lambda \lambda^{-d/2} x^+ \ , \ \ x^-\to {x^-}'=\lambda \lambda^{d/2} x^-,
	 	\ee
	 which is nothing but a combination of a dilatation by a factor $\lambda$ and a boost in the negative $x^p$-direction. Further details on the interpretation of the isometries of $K_{d+1}$ may be found in section 5 of \cite{Brecher:2000pa}.\\
	 \indent As for the matching of global fermionic symmetries, it was shown in the Appendix B of \cite{Cvetic:1998jf} that the metric (\ref{3Kaigorodov}) preserves $1/4$ of the supersymmetry\footnote{As a matter of fact, it can be shown that a pp-wave in AdS described by (\ref{3NHansatz}) always preserves $1/4$ of the supersymmetries, no matter the specific form of $F$ (see also section 2.2 of \cite{Brecher:2000pa} for a detailed derivation).}. Its counterpart on the field theory side now comes naturally from the spontaneous breaking of supersymmetry caused by the non-vanishing vev of the light-front Hamiltonian.
	
	\section{Conclusions}
	In this paper we have presented evidence in favor of the conjecture that the gravity dual to the light-front vacuum state of a CFT quantized on the light-front and in the large $N$, strong coupling limit takes the form of a pp-wave propagating in $AdS$ parallel to the conformal boundary (\ref{3NHansatz}). The simpliest example of such pp-wave spacetimes is the Kaigorodov spacetime $K_{d+1}$ (\ref{3Kaigorodov}), for which the profile of the wave is independent of both $x^+$ and the transverse coordinates $x^a$. In particular, we have computed holographically the (light-front) vacuum expectation value of the CFT energy-momentum tensor that arises from a large class of asymptotically $AdS$ pp-waves, including $K_{d+1}$. In turn, such a vev was shown to be coherent with our discussion of light-front quantization in section 2. We have also discussed how the isometries of $K_{d+1}$ have a natural interpretation as the subgroup of bosonic symmetries of the LFV than remain unbroken by a constant non-vanishing vev of the light-front Hamiltonian in the dual field theory. Finally, given the fact that in the CFT the non-vanishing vev of the light-front Hamiltonian is entirely due to the zero modes, we have suggested that the zero mode sector is implemented holographically through the propagating gravitational wave and that equation (\ref{3NHF}) may be regarded as a strong coupling version of the zero mode constraint equation.

	\section*{Acknowledgments}
	I would like to thank Guy de T\'eramond, Hugo Sol\'is, and Gen\'is Torrents for their valuable suggestions on improving this manuscript. I also want to thank Jos\'e Mariano Gracia,  Miguel Guzm\'an and Alejandro Jenkins for all their support during my firsts days in Tiquicia. This work wouldn't have been possible without their help.

\end{document}